\documentclass[journal=jpcafh,manuscript=article]{achemso}

\usepackage[version=3]{mhchem} 
\usepackage[bookmarks,bookmarksopen]{hyperref}
\usepackage{graphicx}
\usepackage{dcolumn}
\usepackage{bm}

\newcommand{\HH}    {\mbox{H$_2$}}           
\newcommand{\HHD}    {\mbox{HD}}           
\newcommand{\DD}      {\mbox{D$_2$}}         
\newcommand{\NN}    {\mbox{N$_2$}}           
\newcommand{\HHHp}    {\mbox{H$_3^+$}}       
\newcommand{\HHDp}    {\mbox{H$_2$D$^+$}}    
\newcommand{\HDDp}    {\mbox{D$_2$H$^+$}}      
\newcommand{\NNHp}    {\mbox{N$_2$H$^+$}}       
\newcommand{\NNDp}    {\mbox{N$_2$D$^+$}}    
\newcommand{\HHO}  {\mbox{H$_2$O}}         
\newcommand{\HDO}  {\mbox{HDO}}         
\newcommand{\HHCO}  {\mbox{H$_2$CO}}         
\newcommand{\HCN}  {\mbox{HCN}}         
\newcommand{\DCN}  {\mbox{DCN}}         

\newcommand{\CHHHp}   {\mbox{CH$_3^+$}}      
\newcommand{\CHHDp}   {\mbox{CH$_2$D$^+$}}      
\newcommand{\CHDDp}    {\mbox{CHD$_2^+$}}      
\newcommand{\CDDDp}   {\mbox{CD$_3^+$}}      
\newcommand{\CHHHDn}   {\mbox{CH$_{3-n}$D$_n^+$}}      
\newcommand{\CHHHHHp}   {\mbox{CH$_5^+$}}      

\newcommand{\DCOp}  {\mbox{DCO$^{+}$}}       
\newcommand{\HCOp}  {\mbox{HCO$^{+}$}}       
\newcommand{\emm}[1]{\ensuremath{#1}}   
\newcommand{\emr}[1]{\emm{\mathrm{#1}}} 
\newcommand{\unit}[1]{\emm{\, \emr{#1}}}

\newcommand{\pscm}{\unit{cm^{-2}}}

\newcommand{\pcm} {\unit{cm^{-1}}}
\newcommand{\kms}   {\unit{km\,s^{-1}}}

\author{Evelyne Roueff}
\email{evelyne.roueff@obspm.fr}
\affiliation[LUTh and UMR 8102 du CNRS]
{Observatoire de Paris, Place J. Janssen, 92190 Meudon, France}

\author{Maryvonne Gerin}
\affiliation[ENS, LERMA and UMR8112 du CNRS]
{D\'epartement de Physique de l'ENS, 24 rue Lohmond, Paris 75230 cedex 05, France}

\author{Dariusz C. Lis}
\affiliation[Cahill Center for Astronomy and Astrophysics]
{MC301-17, Caltech, Pasadena, Ca91125, USA}

\author{Alwyn Wootten}
\affiliation[North America ALMA Science Center]
{NRAO, 520 Edgemont Rd., Charlottesville, Virginia 22903}

\author{Nuria Marcelino}
\affiliation[NRAO]
{NRAO, 520 Edgemont Rd., Charlottesville, Virginia 22903}

\author{Jose Cernicharo}
\affiliation[CAB. INTA-CSIC]
{28850 Torrejon de Ardoz. Madrid. Spain}

\author{Belen Tercero}
\affiliation[CAB. INTA-CSIC]
{28850 Torrejon de Ardoz. Madrid. Spain}

\title[\texttt{achemso}\CHHDp]
{\CHHDp, the Search for the Holy GRAIL}

\begin{document}
\begin{abstract}
  \CHHDp,  the singly deuterated counterpart of \CHHHp, offers an alternative way to mediate formation of deuterated species 
  at temperatures of several tens of~K, as compared to the release of deuterated
  species from grains. 
  We report a longstanding observational search for this molecular ion, whose rotational spectroscopy is not yet completely secure.
  We summarize the main spectroscopic properties of this molecule  and discuss 
  the chemical network leading to the formation of  {\CHHDp}, with explicit account of the ortho/para forms of  {\HH}, {\HHHp} and \CHHHp. Astrochemical models
  support the presence of this molecular ion in moderately warm environments at a marginal level.  
 \end{abstract}

\section{Introduction}
 {\CHHHp} is recognized as a key polyatomic molecular ion in astrophysical plasmas. It has been found in the innermost coma of comet Halley \cite{balsiger:87,haider:93,rubin:09} and is thought to be present both in diffuse  and dense molecular clouds \cite{indriolo:10}.  
 Despite its importance, there is a lack of high-resolution spectroscopic data, primarily because of its polymerization in discharges \cite{crofton:88}. Laboratory infrared spectroscopy studies have first been conducted in  Oka's group\cite{crofton:88}. More recently, threshold photoionization studies of the methyl radical and its deuterated isotopologues  in the 9.5 - 10.5 eV photon energy range have allowed to investigate the vibrational spectroscopy
 of their corresponding cations by using  the easily tunable and powerful sources of radiation, provided by the third generation of synchrotron sources \cite{alcaraz:10}. Rotational spectroscopy of \CHHHp , however, cannot be achieved as its fully symmetric $D_{3h}$  ground state planar structure does not allow  for a permanent dipole moment. 
 
Deuterium substitution of a hydrogen atom in {\CHHHp}  breaks the symmetry and allows the presence of a small, but significant 0.3 D dipole moment, which gives potentially observable rotational transitions of  {\CHHDp}.
Regrettably, the molecule is light and the rotational constants are large, producing a widely spaced level structure. Two independent recent laboratory studies \cite{amano:10,gartner:10} have  considerably improved our knowledge of the {\CHHDp} rotational spectrum, which is now an entry in the CDMS database (Cologne Database for Molecular Spectroscopy) \cite{mueller:01,CDMS:05} (http://www.astro.uni-koeln.de/cdms/). 

{\CHHHp} does not react  with molecular  nor atomic hydrogen,  as the corresponding reactions are highly endothermic. Nevertheless, molecular hydrogen can radiatively associate in a slow reaction where a temporary molecular complex {\CHHHHHp}  is formed and stabilized through infrared emission. Alternatively, {\CHHHp} can exchange a deuteron with HD, producing {\CHHDp}, and offering an efficient pathway to deuteration, as will be discussed later. It may also react with other abundant neutral molecules, suggesting a natural gas-phase path to molecular complexity \cite{smith:93}.  

 \section{Spectroscopy of  the {\CHHHDn} family}
 \label{sec:prop}
 \subsection{Vibrational and rotational constants}
Values of the {$\nu_3$} fundamental frequency of {\CHHHp}  (ref \cite{crofton:88}), $\nu_1$ fundamental frequency of  {\CHHDp} and  {\CHDDp}  (ref. \cite{jagod:92}), as well as $\nu_4$ frequency of  {\CHHDp}  (ref \cite{jagod:92})  have been derived from the 
infrared vibrational spectrum of these ions performed in the group of Oka. As a four atoms containing species, 6 vibrational modes are involved. For the fully substituted species {\CHHHp}  and  {\CDDDp},  the $\nu_3$ stretching and $\nu_4$  bending modes are doubly degenerate. Threshold photoelectron spectroscopy studies with highly tunable radiation sources provided 
by the new third generation synchrotron facility SOLEIL allow to record the photoionization spectra over a wide range of photoionization energies. All the vibrational frequencies of the methyl radical and its deuterium-substituded forms, as well as those of the corresponding ions have been reported by Cunha de Miranda  {\it{et al.}}, 2010  \cite{alcaraz:10}. Additional theoretical quantum calculations are also documented by these authors \cite{alcaraz:10}.

 \begin{table}[h]
    \caption {Spectroscopic constants of {\CHHHp} and its  deuterated isotopologues. 
      }
    \begin{center}
      \begin{tabular}{ccccc} 
        \hline \hline
          &  {\mbox{CH$_3^+$}} &    {\mbox{CH$_2$D$^+$}}    &  {\mbox{CHD$_2^+$}}      &   {\mbox{CD$_3^+$}}\\
 \hline
        electronic state&  $\tilde{X} ^1A'_1$ &    $\tilde{X} ^1A'_1$   &  $\tilde{X} ^1A'_1$    &   $\tilde{X} ^1A'_1$  \\
point group        &  $D_{3h}$ &   $ C_{2v}$  &   $ C_{2v}$  &    $D_{3h}$  \\
 \hline
\hline
  $\nu_1^0$ (\pcm) &  3037  \cite{keceli:09}   &    &     &         \\
    $\nu_1$ (\pcm) &   2940  \cite{alcaraz:10} &  3005 \cite{jagod:92} &  3056 \cite{jagod:92} &   2097  \cite{alcaraz:10}  \\
 symmetry & $a'_1$ symmetric stretch    &  $a_1$ &  $a_1$&  $a'_1$  \\
\hline
$\nu_2^0$ (\pcm)   &  1418 \cite{keceli:09} &      &           &                      \\
$\nu_2$ (\pcm)    &  1359 $\pm$ 7 \cite{liu:01} &2240 \cite{alcaraz:10} &2168 \cite{alcaraz:10} &  1085 \cite{alcaraz:10} \\
 symmetry  &  $a"_2$  OPLA $^a$&  $A_1$  &  $A_1$ &  $a"_2$    \\
 \hline
$\nu_3^0$ (\pcm)  &  3247 \cite{keceli:09} &  & &     \\
 $\nu_3$ (\pcm) & 3108 \cite{crofton:88}  &  1389 \cite{alcaraz:10} &  1036$^{(2)}$   & 2345 \cite{alcaraz:10}  \\
 symmetry  &   $e'$ degenerate stretch &  $a_1$  &  $a_1$ & $ e' $  degenerate stretch  \\
\hline
$\nu_4^0$ (\pcm)    &  1429 \cite{keceli:09} &  &  &    \\
 $\nu_4$(\pcm) &  1370 \cite{liu:01}  & 1299 \cite{alcaraz:10}   &  1188 \cite{alcaraz:10}  &  1030  \cite{alcaraz:10} \\
   symmetry     &  $e'$ degenerate bend  &   $b_1$ &  $b_1$  &$ e' $ degenerate bend    \\
 \hline
 $\nu_5 $ (\pcm)&   &    3106  \cite{jagod:92} &  2356 \cite{jagod:92} &     \\
  symmetry     &   &   $b_2$ &  $b_2$  &     \\
 \hline
 $\nu_6 $ (\pcm)&      &   1171 \cite{alcaraz:10}  & 1281 \cite{jagod:92} &    \\
  symmetry     &    &   $b_2$ &  $b_2$  &    \\
 \hline
 A (\pcm)   &                             &   9.3686865 \cite{amano:10}  &      7.25251 \cite{jagod:92} &               \\
 B  (\pcm)  &   9.3622 \cite{kraemer:91}  &  5.7713018 \cite{amano:10}  &     4.69046 \cite{jagod:92}   &   4.731\cite{kraemer:91} \\
 C  (\pcm) &    4.7155 \cite{kraemer:91} &  3.5252332 \cite{amano:10}&    2.815470 \cite{jagod:92}   &    2.364 \cite{kraemer:91} \\
 \hline
 \end{tabular}
 \end{center}
 $^a$ OPLA stands for "Out of Plane Large Amplitude" mode\\
 $\nu_i^0$ stands for theoretical harmonic wavenumber.
Experimental values of the fundamental wavenumbers $\nu_i$, when available, are displayed. A, B, C
refer to the rotational constants.
     \label{tab:spec}
  \end{table}

\begin{figure}
    \centering
      \includegraphics[height=1\hsize{},angle=0]{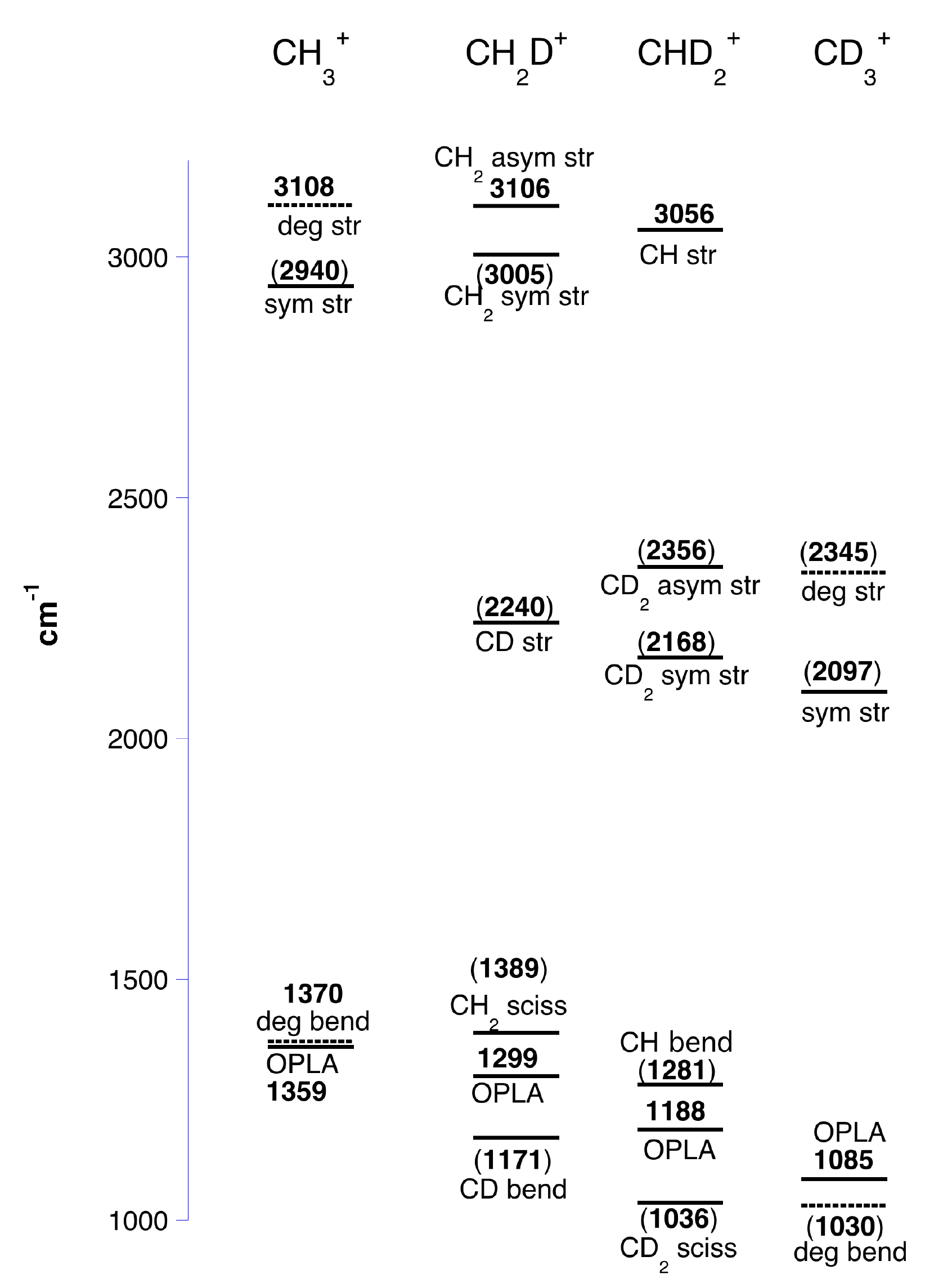}
       \caption{Diagram showing the fundamental vibrations of methyl cation isotopologues. The experimental vibrational frequencies are listed when available. Theoretical values are given in parenthesis. Dotted lines correspond to degenerate modes of vibration. OPLA stands for "Out of Plane Large Amplitude" mode. Adapted from Brum  {\it{et al.}}, 1993  \cite{brum:93}.
                   }
          \label{fig:chnp}
   \end{figure}

We display the spectroscopic constants of the different deuterated isotopologues in Table \ref{tab:spec}. Theoretical harmonic frequencies $\nu_i^0$  of  {\CHHHp} have also been computed by Keceli {\it{et al.}}, 2009  \cite{keceli:09} and are also reported. 
Finally, Table \ref{tab:spec}  also gives  the rotational spectroscopic constants of the vibrational ground state of the different isotopologues.
Figure \ref{fig:chnp}, adapted from  Brum  {\it{et al.}} 1993 \cite{brum:93}, summarizes the respective numerical values of the fundamental vibrational frequencies on a wavenumber scale for all deuterated  isotopologues of the methyl cation. 

\subsection{Zero Point Energies (ZPEs)}
Determining zero point energies (ZPEs) is important for predicting the relative stability of different isotopologues as they can be further used to determine the exothermicity of isotopic exchange reactions. Whereas ZPEs values are  well known and documented for diatomics, as summarized in Irikura, 2007 \cite{irikura:07},
their estimates, computations and experimental corrections are much less obvious for polyatomics (Csonka {\it{et al.}} 2005)\cite{csonka:05}. 
Using a fourth-order expansion of the potential energy, the second order perturbative expression of the vibrational energy of asymmetric tops can be expressed as \cite{barone:04}:
\begin{equation}
E_{n}= \chi_0 +  \sum_i{  g_i  \nu_i^0 (n_i+ \frac{1}{2})}+ \sum_iÊ\sum_{j \prec i}  g_i g_j\chi_{ij} (n_i+\frac{1}{2}) (n_j + \frac{1}{2})    
\end{equation}
where $\nu_i^0$ is the harmonic frequency and the anharmonic constants $\chi_{ij}$ are simple functions of the second, third and semi-diagonal fourth energy derivatives w.r.t. normal modes. One can then derive the expressions of the fundamental vibrational frequencies $\nu_i$ and ZPE :
\begin{equation}
\nu_i = \nu_i^0 + 2 \chi_i + \frac{1}{2} \sum_{j \ne i}  \chi_{ij} 
\end{equation}
and 
\begin{equation}
ZPE=\chi_0 + \frac{1}{2} \sum_{i} ( \nu_i^0 + \frac{1}{2}\chi_{ii} + Ê\sum_{j \prec i} \frac{1}{2}  \chi_{ij})Ê
\end{equation}
Different methods have been proposed to obtain improved ZPE's by combing the formulae including harmonic vibrations and fundamental vibrations.
We follow the recent prescription 
\cite{csonka:05} for computing the ZPE of the methylium molecular ion: 
\begin{equation}
ZPE_{vib}= \sum_i {{  g_i  ( \frac{3}{8}h\nu_i^0 +\frac{1}{8}h\nu_i)}}   
\end{equation}
The ZPEs of the deuterated substitutes of  {\CHHHp} are estimated from the formula including the fundamental frequencies only, as no other information is available:
\begin{equation}
ZPE'_{vib}=  \sum_i {{  g_i  \frac{1}{2} h\nu_i}}   
\end{equation}

 We also compute the corresponding value for \CHHHp. The values are reported in Table \ref{tab:ZPE}. The resulting accuracy is difficult to estimate and the two values displayed for {\CHHHp}, differing by an amount of $\sim$ {69 \pcm},  reflect this issue. 

 {\CHHHp}  has a planar D$_{3h}$ structure and, as such, the ground rotational (ortho) level 0$_{00}$ is forbidden, as in the case of  {\HHHp}, due to Fermi statistics. The first rotational level of  {\CHHHp} is then the para form level 1$_1$. The corresponding rotational energy terms are expressed as :
 \begin{equation}
 E_{J,K}=  BJ(J+1) + (C-B)K^2
 \end{equation}
  With the values given in Table \ref{tab:spec}, the first energy terms of level 1$_1$ (para) and  1$_0$ (ortho) are respectively 14.08  and 18.72  \pcm, measured from the ground vibrational term.
This additional rotational energy term is quoted explicitly for para and ortho forms of {\CHHHp}. 
We also display the ZPE values of \HH, \HHO, HCN and deuterated isotopologues which will be used further to determine the energy released in possible deuterium exchange reactions. These values include the anharmonic contribution.

 \begin{table}[h]
    \caption {Zero Point Energies in \pcm.
  }
    \begin{center}
      \begin{tabular}{lccccc} 
 \hline
          &  p-{\mbox{CH$_3^+$}} &    o-{\mbox{CH$_3^+$}} & {\mbox{CH$_2$D$^+$}}    &  {\mbox{CHD$_2^+$}}      &   {\mbox{CD$_3^+$}}\\
        \hline \hline
 ZPE  (Eq. 4)  & 6834.5 + 14.08 & 6834.5 + 18.72    &  &          &      \\
 ZPE'  (Eq. 5)   &  6903.5  + 14.08 &   6903.5 +18.72  & 6105   &     5542.5     &   4966      \\
  \hline \hline
   &   &   &   &   \\
 \end{tabular}
 
 \begin{tabular}{lccccccc} 
\hline
 &   \HH& {\mbox{HD}}  & \DD&  \HHO  &     \HDO   &   \HCN  &   \DCN \\
  \hline

ZPE &  2179.3 \cite{irikura:07}& 1890.3 \cite{irikura:07}& 1546.5 \cite{irikura:07}& 4638.3\cite{hewitt:05} & 4022.8\cite{hewitt:05}  &   3479.2 \cite{mellau:08} & 2883.9 \cite{moellmann:02}\\
   \hline
\hline
\end{tabular}
 \end{center}
     \label{tab:ZPE}
  \end{table}
\subsection{Fractionation reactions}
\label{sec:frac}

 Significant enhancements of deuterated molecules compared to the 
elemental D/H ratio of about 1.5 $\times$ 10$^{-5}$ have been found  towards cold,
dense and CO-depleted molecular cores 
\cite{gerin:06,daniel:07,caselli:08,roueff:05}. 
Indeed, the observed {\HCOp}/{\DCOp} and  {\NNDp}/{\NNHp} ratios
 reach high values only in the coldest clouds
 and the deuterated variants are practically unobservable in a warm cloud such as OMC-1 
\cite{wootten:82}, in fair agreement with gas-phase chemistry predictions. 
\DCOp~ and \NNDp~are essentially formed from the reaction of \HHDp~with CO  and \NN, so that the
{\DCOp}/{\HCOp}  and~\NNDp / \NNHp ratios reflect essentially the {\HHDp} / {\HHHp} ratio, which decreases
rapidly with temperature, as deuterium is loosely bound in the \HHDp ~molecular ion.
 The exothermicity of the formation
reaction via \HHHp~+ HD has a value of about 230 K and the presence of
ortho-\HH ~reduces drastically the barrier of the backward reaction as
discussed in Pagani {\it{et al.}}, 2009 \cite{pagani:09}. 

It has been recognized quite early that deuterium enrichment could also proceed via fractionation reactions of {\CHHHp} with HD \cite{watson:76}. The exothermicity of this reaction  was then estimated to be of the order of 300 K.  Smith and Adams \cite{saa:82} subsequently derived a value of 370 K from an experimental study at low temperatures. This same value of 370 K has been used in later studies for reactions between {\CHDDp} and HD, when multiply deuterated molecules have been introduced in chemical models \cite{roberts:04}, following their first detection in the interstellar medium \cite{roueff:00,loinard:00}.  
We revisit the possible fractionation reactions involving {\CHHHDn}  in the light of recent ion-molecule studies \cite{gerlich:02b,asvany:04} and
selection rules in ortho/para transitions summarized in Oka, 2004 \cite{oka:04},  involving conservation of the total nuclear spin. 
We display in Table \ref{tab:reac} the different fractionation reactions and measured reaction rate coefficients at two different temperatures. We also report our computed exothermicities $\Delta E$, obtained from the previously determined ZPEs (Table \ref{tab:ZPE})
using fundamental vibrational frequencies: For an exothermic  reaction $R_1 +  R_2 \to  P_1  + P_2$, the corresponding exothermicity is given by:
\begin{equation}
\Delta E = ZPE(R1)+ZPE(R2)-ZPE(P1)-ZPE(P2)
\end{equation}
The exothermicities involved in {\CHHHDn} reactions  with HD are typically larger than 300 K ~ as previously reported,  and we find that the first reaction of  {\CHHHp} with HD  even has an exothermicity close to 650 K, about a factor of 2 larger than the previous estimates, increasing the formation efficiency of {\CHHDp} in warm conditions. The main uncertainty arises from the spectroscopic constants of \CHHDp, where only fundamental vibration frequencies have been derived. Similar values are obtained  in Parise et al., 2009 \cite{parise:09} based  on ZPE values   from  older theoretical values of vibrational frequencies reported in DeFrees \& McLean, 1985 \cite{defrees:85}. 

\begin{table*}
  \caption{Rate coefficients for various deuterium exchange reactions and exothermicities (computed from ZPEs) of the forward reaction. } \label{tab:reac}
    \begin{center}
      \begin{tabular}{lclccc} 
 \hline
          &    &    & exothermicity & \multicolumn{2}{c}{rate coefficient of the  forward reaction} \\
                         &   reaction  &    &               (K)   &     (cm$^3$ s$^{-1}$) &   (cm$^3$ s$^{-1}$)   \\
                        &    &     &                         &     15K                              &     80K     \\
       \hline \hline

\CHHHp  + HD &$\rightleftharpoons$& \CHHDp + \HH   & 654 &  1.65$\pm$ 0.1~(-9) \cite{asvany:04} $^a$ &   \\
\CHHHp (para) + HD &$\rightleftharpoons$& \CHHDp + \HH (para)  & 654 &  4.0~(-10) \cite{asvany:04} $^b$ &   \\
\CHHHp (para) + HD &$\rightleftharpoons$& \CHHDp + \HH (ortho)  & 483 &  6.7~(-10) $^c$  &  \\
\CHHHp (ortho) + HD &$\rightleftharpoons$& \CHHDp + \HH (para)  & 660 &  1.9~(-10) $^c$  &  \\
\CHHHp (ortho) + HD &$\rightleftharpoons$& \CHHDp + \HH (ortho)  & 489 &  1.3~(-9) $^c$ &  \\
\CHHHp + HD &$\rightleftharpoons$& \CHHDp + \HH    &    654 &     & 1.1 (-9) \cite{saa:82} \\
\CHHDp + HD &$\rightleftharpoons$ &\CHDDp + \HH &  393   &   1.59 $\pm$ 0.1 (-9)   \cite{asvany:04} &  7.4 (-10) \cite{saa:82}\\
\CHDDp + HD &$\rightleftharpoons$ &\CDDDp + \HH & 414  &  1.50 $\pm$ 0.1  (-9)  \cite{asvany:04} &   6 (-10) \cite{saa:82}\\
\CHHHp + \DD &$\rightleftharpoons$ &\CHDDp + \HH &   969 &     &     6.6 (-10) \cite{saa:82}\\
\CHHHp + \DD &$\rightleftharpoons$ &\CHHDp + HD &   575&  & 4.4 (-10) \cite{saa:82} \\
\CHHDp + \DD &$\rightleftharpoons$& \CDDDp + \HH &   728 &   &  3 10(-10) \cite{saa:82} \\
\CHHDp + \DD &$\rightleftharpoons$ &\CHDDp + HD &   315 &  & 9 (-10) \cite{saa:82}\\
\CHDDp + \DD &$\rightleftharpoons$ &\CDDDp + HD &   335 &   &  7.4 (-10) \cite{saa:82}\\
\hline
\CHHHp + \HDO &$\rightleftharpoons$& \CHHDp + \HHO &   178 &  
                &    \\
\CHHHp + \DCN &$\rightleftharpoons$& \CHHDp + \HCN  &   207 & 
 &  \\
\hline
    \end{tabular}
  \end{center}
   Values in parentheses refer to power of 10. \\ $^a$ : He buffer gas, pure HD target \\
  $^b$ : para-{\CHHHp}  in para-{\HH}  buffer gas \\
  $^c$: present derivation from branching ratios displayed in Table \ref{tab:exch}. See text.
 \end{table*}

The measurements by Asvany et al. \cite{asvany:04} showed that the reaction rate coefficient of  para-{\CHHHp} in para-{\HH} 
buffer gas is much smaller than the value obtained with a He buffer gas involving pure HD target. We introduce the reactions between {\CHHHp} and HD involving specific para/ortho forms of  {\CHHHp} and {\HH} and derive the corresponding rate coefficients by including the expected branching ratios obtained from the selection rules given in Oka, 2004 \cite{oka:04}.
The corresponding values are summarized in Table~\ref{tab:exch} and differ  slightly from those previously assumed by Walmsley {\it{et al.}}, 2004 \cite{walmsley:04} for the {\HHHp} + HD reaction. 
\begin{table*}
  \caption{Branching ratio of the \CHHHp + HD $\rightarrow$ \CHHDp + {\HH}  reaction.}\label{tab:exch}
     \begin{center}
      \begin{tabular}{ccccc}
      \hline
                                 &   (o-\CHHDp, o-\HH)   &   (o-\CHHDp, p-\HH)  &    (p-\CHHDp, o-\HH)  &     (p-\CHHDp, p-\HH)       \\
                                 \hline \hline
  p-\CHHHp + HD  &  3/8             &       1/4            &      1/4       &         1/8        \\
  o-\CHHHp + HD   &          6/8     &      1/8            &     1/8        &  0    \\     
  \hline
    \end{tabular}
    \end{center}
    \end{table*}

As an example, the derived rate coefficient of the p-{\CHHHp} in o-{\HH} is 5/3  that of p-{\CHHHp} in p-\HH,
as we do not discriminate the ortho/para forms of \CHHDp. 
The other values are obtained by using the total rate coefficient measured in pure HD and assuming that para and ortho \CHHHp are
in equal proportions in the experiment. Then, 
$k_{(o-\CHHHp + \HHD  \to \CHHDp + o-\HH)}$  = $\frac{7\times 4}{22} \times 1.05\times 10^{-9} $ cm$^3$ s$^{-1}$ and
$k_{(o-\CHHHp + \HHD  \to \CHHDp + p-\HH)}$  = $ \frac{4}{22} \times 1.05\times 10^{-9} $ cm$^3$ s$^{-1}$.

As it has been found that {\CHHHp} does not react with {\HHO},   nor with HCN \cite{anicich:03}, we consider possible deuterium exchange reactions of {\CHHHp} with HDO and DCN, which are also displayed in Table \ref{tab:reac}. The deuterium exchange reaction between {\CHHHp}  and HDO is found exothermic whereas \citeauthor{millar:89} \cite{millar:89}  quote it to be endothermic by a similar amount. We also find that the reaction between {\CHHHp} and DCN  is exothermic. These last reactions have not been studied in the laboratory, but they may contribute to the general balance of deuterium exchange reactions in moderately warm environments. 

\section{Observations}
\label{sec:obs}

 \begin{figure}
   \centering
    \includegraphics[height=0.5\hsize{},angle=0]{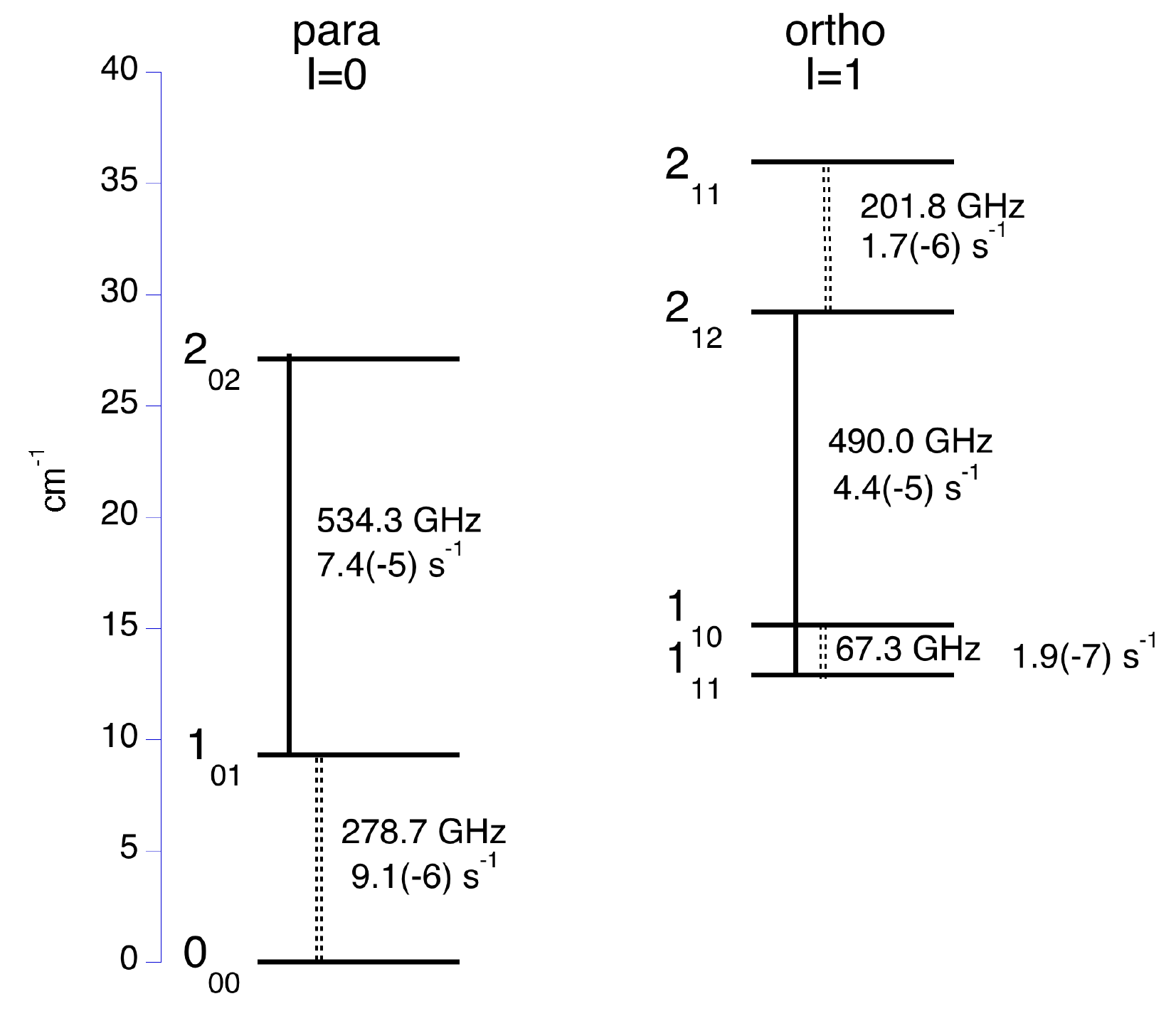}
         \caption{Energy diagram of \CHHDp. Rotational frequencies are listed as well as Einstein coefficients. Full lines refer to laboratory measured transitions \cite{amano:10}. Double dotted lines correspond to theoretical predictions derived
         from fitting the experimental available spectrum with model Hamiltonians. x(y) stands for $x \times 10^y$.}
          \label{fig:chhdp}
  \end{figure}
{\CHHDp} has several transitions which can be studied from the ground and Figure \ref{fig:chhdp} lists the first energy levels and the corresponding transitions as reported in the CDMS catalog. 
Searches of the 490.0 GHz fundamental ortho transition of CH$_2$D$^+$ were performed in May 2007 
using the Caltech Submillimeter Observatory (CSO) on Mauna Kea, Hawaii. No positive detections were 
obtainned towards the low-mass protostar IRAS 16293-2422 and the prestellar core IRAS 16293-2422E, with total 
on source integration times of 70 and 39 minutes respectively.
The corresponding receiver
was 
no longer available during the subsequent observing runs.

Then, spectroscopic observations of the 278.7 and 201.7~GHz rotational lines
of CH$_2$D$^+$ in Orion IRc2 (RA(2000) = 05:35:14.2, DEC(2000)=-5:22:36,
v$_{LSR}$ = 9 \kms)  were  carried out in 2009
January--February using the 230~GHz facility receiver and
spectrometers of the CSO. The weather conditions were good to average, characterized by a
225~GHz zenith opacity of $\sim 0.06-0.12$. The CSO beam size at the
two frequencies is $\sim 26^{\prime\prime}$ and $36^{\prime\prime}$,
respectively, and the main beam efficiency, as determined from total
power observations of Saturn, was $\sim 65\%$ and  {70\%} 
respectively. The 278.7~GHz spectra were taken in the
position-switching mode, with the reference position 10 arc min  away in right
ascension. The velocity resolution
of the FFT spectrometer at the frequencies studied here is about
0.066 and 0.09~\kms, respectively. Pointing was determined by frequent
total power observations of the dust continuum in Orion IRc2.
Additional spectroscopic observations have been performed with the  IRAM   30m telescope, in which we recovered the 
same spectral features as those detected with the CSO. Observations 
were performed in July 2009 (201 GHz) and in February and November 2010 (278 GHz and DCN J=2-1), 
with variable weather conditions at high spectral resolution. 
We used the Eight MIxer Receivers (EMIR) together with the VErsatile SPectrometer Array (VESPA), 
providing 80 kHz of spectral resolution (0.12 and 0.08 km s$^{-1}$ at 201 and 278 GHz, respectively).
All the observations were taken in the Wobbler switching mode, with a secondary throw of 4'. Beam sizes 
at 201 and 278 GHz are $\sim$11'' and 9'' respectively, corresponding to main beam efficiencies of 
$\sim$57\% and 46\%. 
    \begin{figure}
   \centering
    \includegraphics[height=1\hsize{},width=1\hsize{},angle=0]{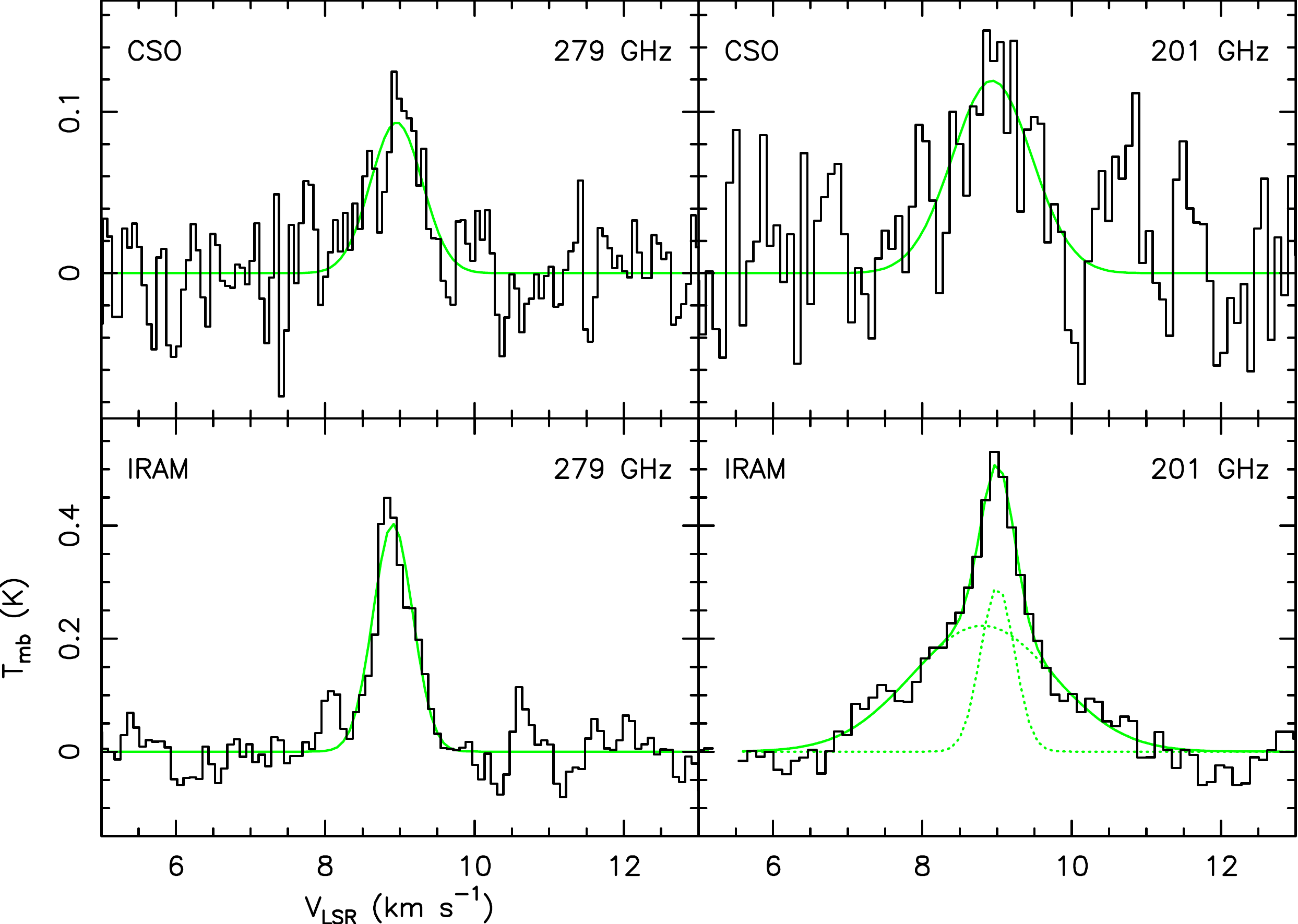}
       \caption{Spectra at the frequencies of the 279~GHz and 201~GHz CH$_2$D$^+$ lines
   toward Orion IRc2. The upper (lower) panels
  show the spectra obtained at CSO (IRAM), where baselines have been removed. 
  The green lines display the corresponding Gaussian
     fits. 
                    }
          \label{fig:obs}
   \end{figure}

Corresponding spectra at  279~GHz and 201~GHz taken at the central (0,0) position are shown in Figure \ref{fig:obs},
where the baselines have been subtracted. 
The first row displays spectra obtained from CSO. 
The total on-source integration time is 125 minutes, with an average
system temperature of 330~K for the 279 GHz transition. The LO frequency has been optimized to
position the CH$_2$D$^+$ line in-between strong lines from the image
sideband. The 201.7~GHz frequency is close to the lower end of the
tuning range of the receiver. The system was  therefore less stable
and the spectra at this frequency were taken in the beam-switching
mode with a secondary chopper throw of 4 arc min.  The  average system temperature is then 525~K
and the corresponding
integration time is 122 minutes.
The second row displays the IRAM observations. Total on-source integration times are 5.8~hrs for both
transitions with system temperatures ranging between 160 to 400 K and opacities of 0.06--0.3
at 279~GHz.  
The  parameters obtained from the gaussian fits are reported in Table \ref{tab:obs} and are remarkably consistent 
in terms of line position and width between the two telescopes. 
 \begin{table}[h]
    \caption {Parameters derived from the CSO and IRAM spectroscopic observations.}
\begin{center}
      \begin{tabular}{cccccc} 
        \hline \hline
         Frequency & Transition & Telescope & Velocity & width & Area    \\
               MHz                &   &      &    km/s      & km/s & K km/s          \\
         \hline
     278691.8 $^{(1)}$ &   1$_{01}$ - 0$_{00}$ (para)       &  CSO  &  8.946   &  0.748 &   0.073  \\
                             &                                      &  IRAM &  8.909  & 0.636 &   0.269 \\
    201754.2  $^{(2)}$ &   2$_{11}$ - 2$_{12}$ (ortho)   &  CSO & 8.933   &  1.25  &   0.159 \\
                                                 &                    &  IRAM & 9.007    &  0.534 &   0.166 \\
                                             &                  &  IRAM &      8.803   &  2.23 &   0.530 \\
        \hline
         \label{tab:obs}
         \end{tabular}
  \end{center}
 $^{(1)}$  the corresponding frequency reported by Amano \cite{amano:10} is 278691.656 (26) MHz.\\
  $^{(2)}$  the corresponding frequency reported by Amano \cite{amano:10} is  201753.947(70) MHz.
 \end{table}      
We also report the frequency values derived by Amano \cite{amano:10} who was able to detect  pure rotational transitions at high frequencies, providing strong constraints on the subtle distortion constants involved in the model Hamiltonian. These values  differ by  less than 100 kHz from  those reported in CDMS, which  are deduced from high spectral resolution laser induced infrared spectra \cite{gartner:10,gartner:13}. The frequency agreement between our "observed" value
and these predictions is excellent.
Thanks to the higher spatial resolution available at IRAM and the higher signal to noise ratio, the 201 GHz feature  may 
consist of 2 velocity components, the narrower one consistent with the value derived for the 278 GHz transition. 
Indeed, the Orion-IRc2 source is heavily congested with lines and great care is required before claiming a detection. The 201 GHz range is particularly 
 delicate as several methyl formate (E species) transitions occur at a very close frequency (201.75325 GHz), however originating from quite excited levels (E $\sim$ 300K). Si$^{18}$O (J=5-4) at 201.7515 is also present \cite{tercero:11} and disentangling the various components is beyond the present study.  

%
  \begin{figure}[h]
    \centering
    \includegraphics[width=14.cm]{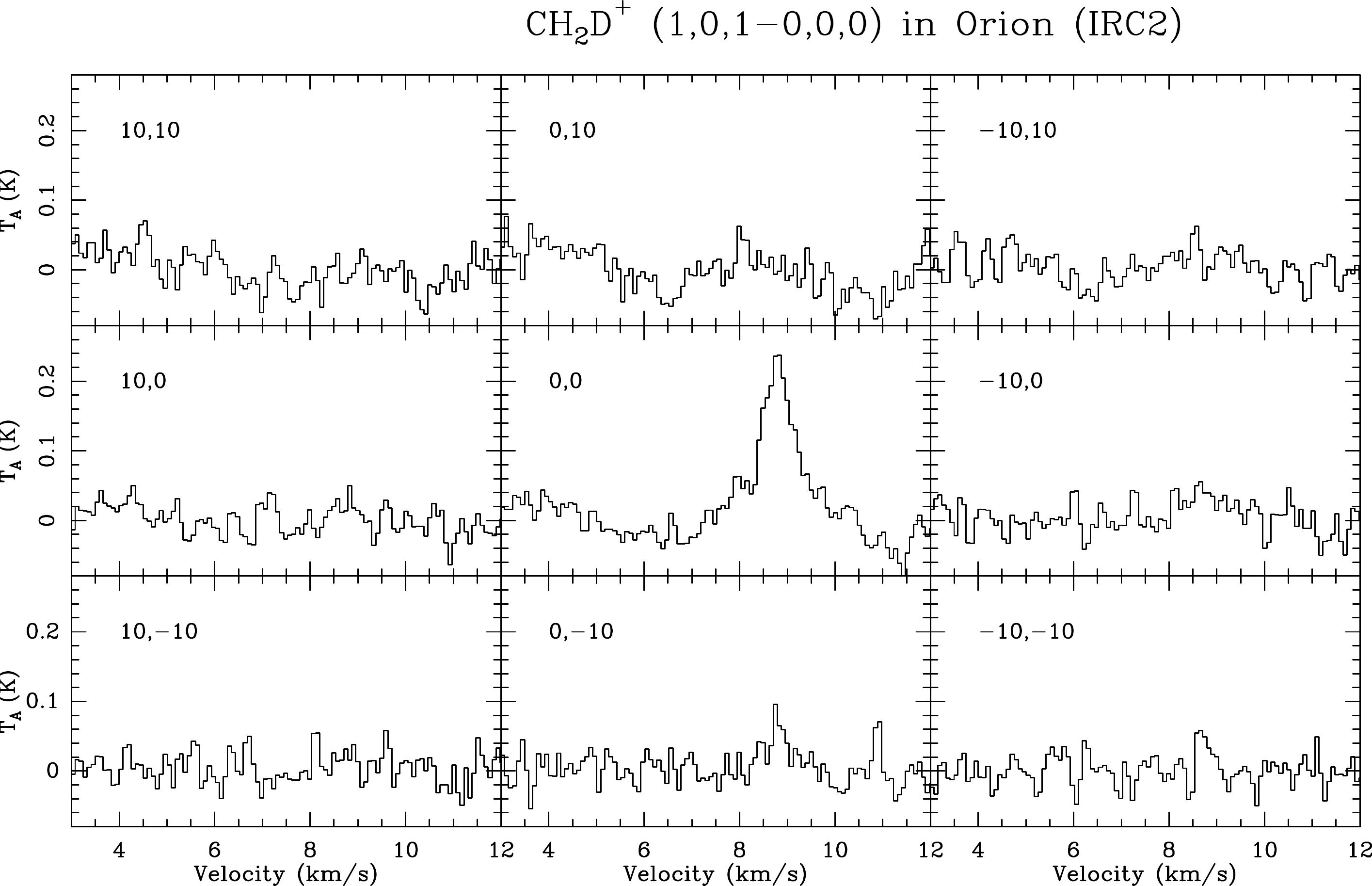}
          \caption{map at 278.7~GHz CH$_2$D$^+$ 1$_{01}$-0$_{00}$ transition
     toward Orion IRc2. 
                    }
          \label{fig:map}
    \end{figure}
    \begin{figure}[h]
    \centering
     \includegraphics[width=14.cm]{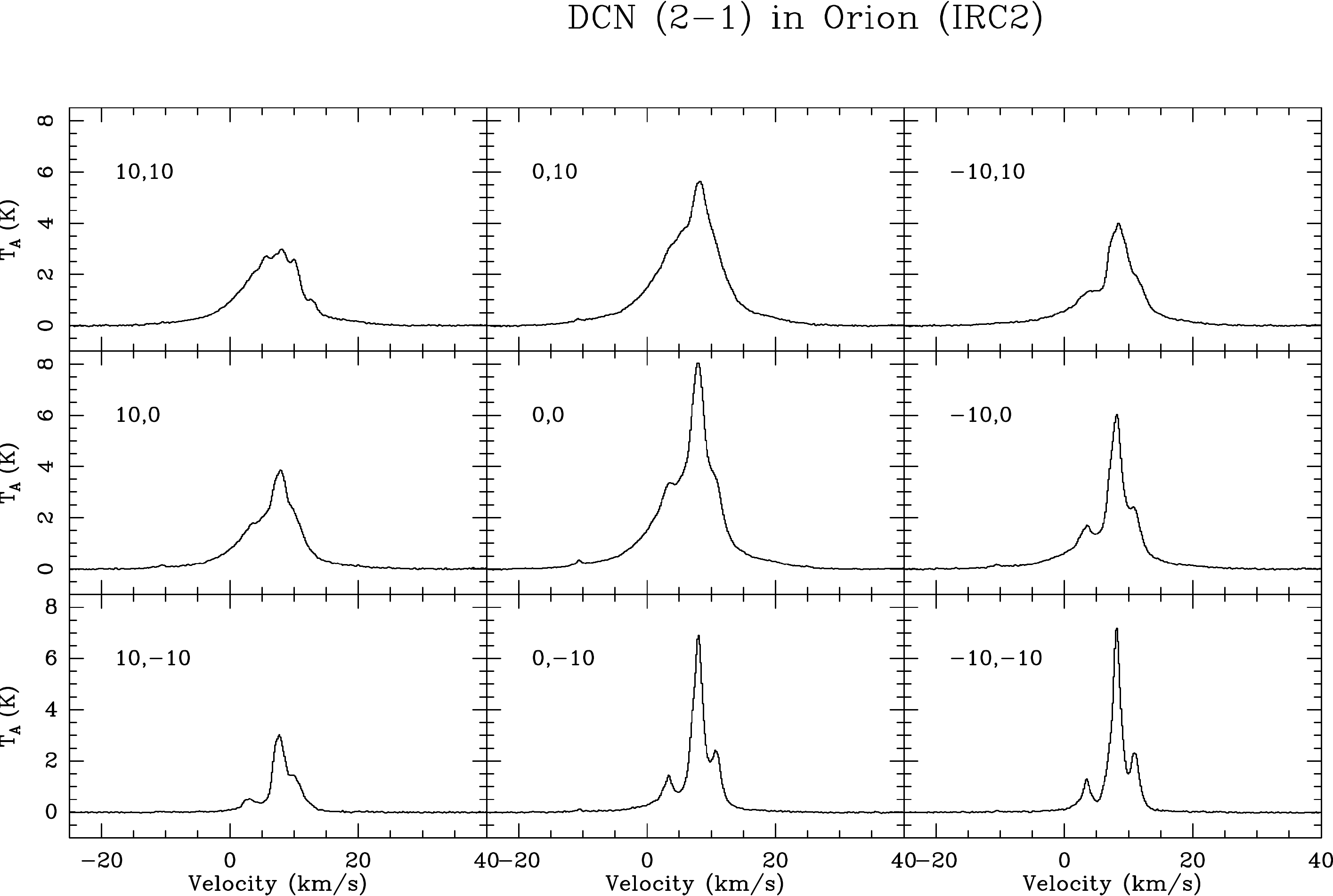}
          \caption{map at 144.8~GHz DCN J=(2-1) transition
     toward Orion IRc2. 
                   }
          \label{fig:mapdcn}
   \end{figure}
We also performed a map of both 201 and 278 GHz 
spectral features around the (0,0) position, together with emission from HDCO and DCN with the IRAM 30m telescope. 
As pointed out earlier, the feature at 201 GHz could be blended with methyl formate emission, thus we 
only display the maps corresponding to the 278 GHz feature in Figure \ref{fig:map}. Total integration times are 
5.8 hrs towards the centre, and $\sim$1.5 hrs at the other positions of the map. Figure \ref{fig:mapdcn} presents emission from DCN J=2-1 observed simultaneously. Both maps show a peak 
at the central position but the 278 GHz feature is much more localized. We thus tentatively conclude from our observations
that the carrier of the detected transitions is \CHHDp and occupies a very limited region. A FWHM  source size of   11.5" is required to reproduce the IRAM and CSO spectra at 278 GHz.
We explore the possible corresponding observational parameters for other  transitions detectable with GBT and Herschel HIFI. 
Table \ref{tab:prediction} displays the  predictions for the corresponding integrated intensities for different values of the excitation temperature and assuming an ortho to para ratio of 3, as expected in the high temperature limit.  The predicted  line intensities for the ortho 201 GHz transition are significantly smaller than the observed ones, which suggests a possible blend.  A full account of these spectral features leads to an  unrealistically high ortho to para ratio and we thus principally aim at reproducing the para 278~GHz  observational parameters, as shown in Table \ref{tab:prediction}.
The observed spectra are modeled using the myXCLASS software package {\footnote{available at  http://www.astro.uni-koeln.de/projects/schilke/XCLASS}} and the
 column densities  are   reported in Table \ref{tab:prediction}. 
\begin{table}[h]
    \caption { Integrated intensities of different  {\CHHDp} transitions in K km/s :  model predictions  
    and corresponding column densities. Observed values are also reported when available.}
\begin{center}
      \begin{tabular}{cccccc} 
        \hline \hline
               &               &    \multicolumn{3}{c}{LTE model}   & Observed   \\
  Telescope   &  Transition &     T= {20 K}   &   T=  {30 K} &   T=   {40 K} &     \\
                          &          MHz  &  K km/s  & K km/s &  K km/s &  K km/s   \\
         \hline
       GBT & 67273.574   &  0.172 &  0.178   &   0.197&     \\
      IRAM & 201754.2   &  0.061  &   0.102 &  0.143 &   0.166 \\
       CSO & 201754.2   &  0.012 &   0.02 &   0.029 &  0.159 \\
     IRAM & 278691.8   &  0.271 & 0.270   &  0.273 &   0.271 \\
       CSO & 278691.8   & 0.071 &0.071   &  0.072 &  0.072\\
       HIFI & 490012.247   &  0.059 & 0.084   & 0.109 &        \\
     HIFI & 534280.117  &  0.033  & 0.051     &  0.064  &   \\
     \hline
     N (ortho)  &($\times$ 10$^{13}$cm$^{-2}$) & 5.67 &    7.5    &     10.5 &    \\ 
     N (para)  &  ($\times$ 10$^{13}$cm$^{-2}$) & 1.89 &      2.5  &  3.35 &   \\ 
     N (total) & ($\times$ 10$^{13}$cm$^{-2}$) & 7.6  &     10.0  &  13.9  &   \\ 
   \hline
         \label{tab:prediction}
         \end{tabular}
  \end{center}
  \end{table}       
Lis et al.\cite{lis:98} have mapped the Orion A Molecular cloud at 350 $\mu$m and obtained a maximum flux of 1300 Jy within a 12"    beam. With a dust temperature of 55 K and an absorption coefficient of 10 cm$^2$/g, the resulting  \HH ~column density is 1.35 $\times$ 10$^{24}$  \pscm. 
Then, the resulting fractional abundance of {\CHHDp} relative to molecular hydrogen is  between 5.6 $\times$ 10$^{-11}$
and 1.0 $\times$ 10$^{-10}$. The derived value is  based on the 278 GHz para transition. 

\section{Astrochemical models}
\label{sec:chim}
Astrochemical models, in which one solves the coupled differential equations describing the time evolution of the various chemical 
species resulting from formation/destruction chemical processes at one given temperature and density,  allow to
check the relevance of the proposed identification. 
To address this point, we amended the chemical network   
used in Pagani {\it{et al.}}, 2011, 2012  \cite{pagani:11,pagani:12},  where ortho/para forms of  \HH, \HH$^+$, {\HHHp} and all their deuterated substitutes are considered separately in the chemical reactions. We introduced para/ortho forms of {\CHHHp} in order to take into account their different behavior in the presence of HD, as discussed previously and shown in Table \ref{tab:reac}. The effect of the small energy difference between the two lowest para and ortho levels of {\CHHHp} (4.6 {\pcm} $\sim$ 7 K) is, however, negligible in comparison with the exothermicities involved in the reactions. This differs from ortho-hydrogen, which has an energy of 118.5 {\pcm} ($\sim$ 170.5 K) above the para level and can significantly reduce the deuterium fractionation efficiency in the   \HHHp  + HD  $\rightleftharpoons$  \HHDp + {\HH}    reaction, as first pointed out by Pagani  {\it{et al.}} \cite{pagani:92}. We included the recent values of the dissociative recombination rates and branching ratios
of {\CHHHp}, as measured by Thomas {\it{et al.}} \cite{thomas:12} in the heavy storage ring CRYRING in Stockholm. Reactions involving nitrogen have  been studied recently \cite{daranlot:12} as advertised in the KIDA database {\footnote{available at http://kida.obs.u-bordeaux1.fr/}} and the corresponding rate coefficients have been modified accordingly.
Then, the chemical network includes 220 atomic and molecular species (where para(ortho) form is counted as a molecular species) and 4024 chemical reactions.

\begin{figure}
   \centering
     \includegraphics[height=1\hsize{},angle=0]{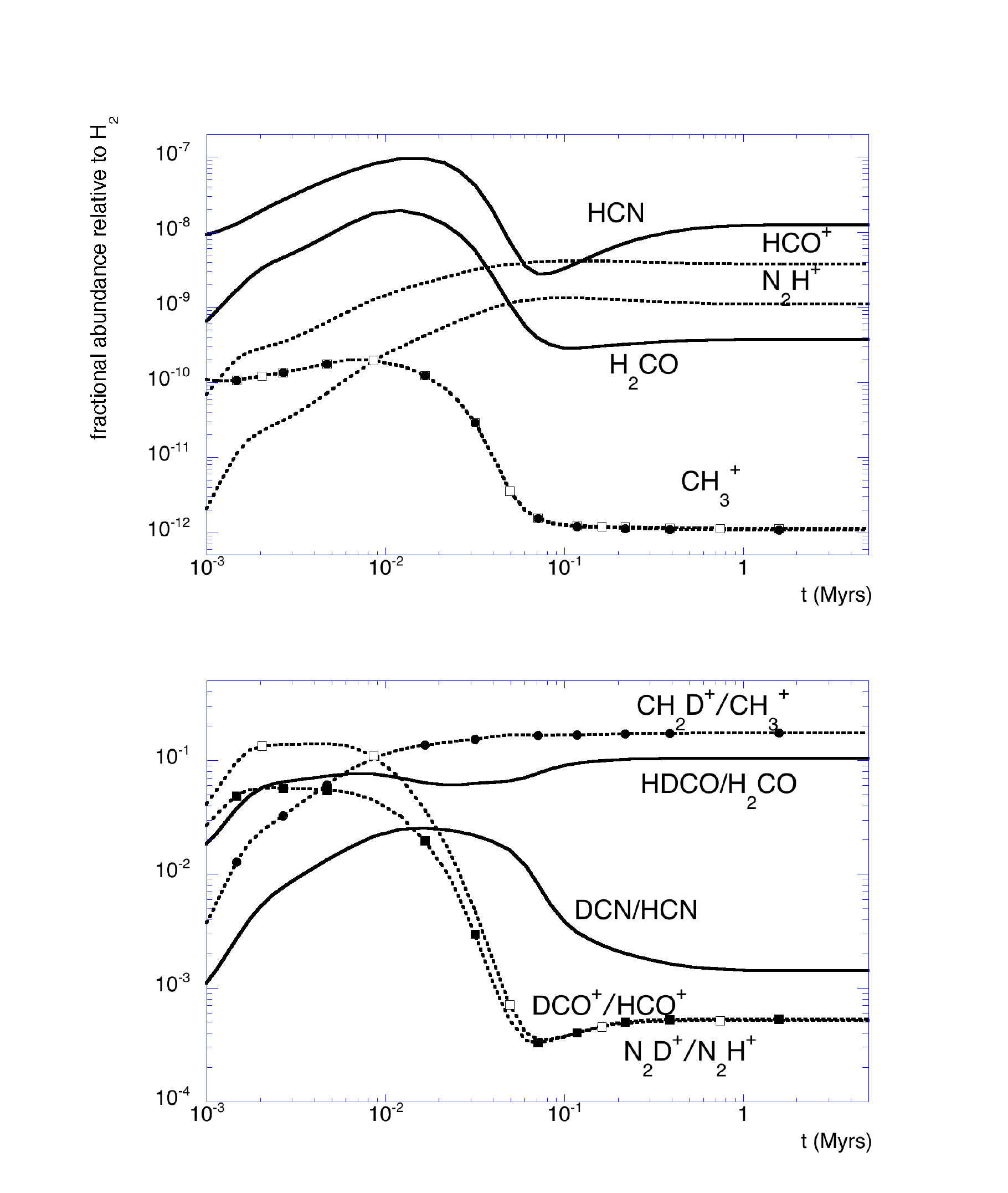}
          \caption{Time evolution results corresponding to   T = 50 K and n(\HH) = 10$^5$ cm$^{-3}$. \\
          Upper panel: time evolution of fractional abundances of \HHCO, HCN, {\CHHHp} (para (open squares) and ortho (full circles)),  {\HCOp} and \NNHp. Neutrals are displayed as full lines, ions as dotted lines. \\
          Lower panel : corresponding deuterium fractionation ratios.
 full circles: \CHHDp/ \CHHHp; open squares:    \DCOp/ \HCOp; full squares: \NNHp /  \NNDp.    }
                   \label{fig:TD}
   \end{figure}

Figure \ref{fig:TD}  displays the time evolution of  selected fractional molecular abundances,  as well as the corresponding deuterium fractionation ratios for T = 50K and a molecular hydrogen density of 10$^5$ cm$^{-3}$. We find that the time evolution profiles of {\HHCO} and {\CHHHp} exhibit a maximum around several 10$^4$ years, the so-called "early time" value \cite{herbst:89}, for which the value of the total {\CHHHp} fractional abundance is about 4 $\times$ 10$^{-10}$, about two orders of magnitude higher than the steady state value, which is reached when the molecular fractional abundance becomes independent of the time scale. We have verified that the location of the maxima and the corresponding fractional abundance values are similar, regardless of the initial condition 
(initial atomic or molecular hydrogen). The steady state values are reached at different times, 
depending on the considered molecules and are found to be independent on the initial conditions, as expected. The deuterium fractionation ratios at the early time peak are also significantly higher than the values at steady state, except for {\HHCO} and {\CHHHp} where the ratios exhibit a small increase from the early time values  up to  steady state. 

In order to test the density and temperature dependences, we display steady state results for density values n(\HH) =  10$^4$ and 10$^5$ cm$^{-3}$, at 3 different temperatures corresponding to medium warm environments in Table \ref{tab:res}, in the range of physical conditions thought to be present in the Orion IRc2 region. 
The steady state ortho-to-para ratio of {\HH} is found to be slightly smaller, but close to its temperature equilibrium value, labelled as ETL. Ortho and para {\CHHHp}  have very similar fractional abundances. 
The deuterium ratio in {\CHHHp} is of the order of 10 $\%$,  even at T = 60 K; however, the fractional abundances are quite small and  close to a detectable limit. The values decrease with increasing densities as well, following
the trend of the fractional ionization.
{\CHHDp}  may also be used to transfer deuterium to HCN, \HHCO,  {\HCOp}, ... via its reaction with atomic oxygen and nitrogen, as emphasized in Roueff  {\it{et al.}}, 2007 \cite{roueff:07}, Parise {\it{et al.}},  2009 \cite{parise:09} and offers an alternative way of deuteration via gas phase processes in medium warm environments. 
As expected, the ratio
of deuterated molecules compared to the main isotopologue is a decreasing function of temperature. The deuteration ratios of HCN,  {\HCOp} and {\NNHp}  are in the 10$^{-3}$ - 10$^{-4}$ range,  whereas that of {\HHCO} is several percent. However, as the fractional abundance of HCN is about fifty times higher than that of \HHCO, even a smaller deuterium fractionation is observable, as shown in the DCN map displayed in Figure~\ref{fig:mapdcn}.
These results can support a possible detection of the {\CHHDp} molecular ion under quite specific conditions: as the fractional abundances are small,  a large column density of molecular hydrogen is required, a range of temperatures between 40 - 60 K  is adequate, and evolution times of several ten thousands years are propitious. Such requirements may be met  
in Orion IRc2.

\section{Conclusions}
We have presented a careful analysis of the gas phase chemistry associated with  {\CHHHp} and its deuterated variants.
ZPEs of the various deuterated isotopologues have been derived in the light of recent spectroscopy experiments and quantum chemical calculations. The role of anharmonicity effects is only included for {\CHHHp},  for which information is available.
The significant exothermicities involved in the deuterium exchange reactions with HD allow an efficient deuterium enrichment  of \CHHHp. However, as {\CHHHp}  may also  react with {\HH} in a slow (but not  insignificant and not well known) radiative association reaction, its fractional abundance remains low,  but exhibits a suggestive maximum at the early time peak during the time evolution. 
We have reported our observational searches of {\CHHDp} in the Orion IRc2 region. Two spectral features detected at 201.7542 and 278.6918 GHz respectively, may correspond to the {\CHHDp} molecular ion. Those values are very close to the frequencies  predicted by  Amano \cite{amano:10}. However, the feature at  201 GHz may be blended with methyl formate and Si$^{18}$O \cite{tercero:11}. 
The feature at 278 GHz is cleaner. However, one single spectral feature is obviously not adequate to claim  a secure identification. 
We also predict integrated intensities for other transitions which may become accessible in the near future.
If real, the {\CHHDp} source in Orion is extremely localized and should be at relatively moderate temperatures below $\sim$ 50K, given the reported line width of less than 1 km s$^{-1}$. In addition,  chemical modeling can positively corroborate such a detection, more favourably when time dependent effects are invoked.  
In spite of the observational challenges,  the updated chemical analysis performed here should help understanding the formation/destruction mechanisms of this puzzling molecular ion and constrain the possible role of gas phase chemistry in the deuteration of interstellar molecules in warm environments.

\begin{table*}
  \caption{Steady state fractional abundances relative to {\HH}  for different values of n(\HH)  and temperature. 
  } \label{tab:res}
    \begin{center}
     \begin{tabular}{l|ll|ll|ll|} 
 \hline
 Temperature          &   \multicolumn{2}{c}{40 K}  &    \multicolumn{2}{c}{50 K}     &    \multicolumn{2}{c}{60 K}  \\
n(\HH) cm$^{-3}$ &   10$^4$  & 10$^5$ &   10$^4$ & 10$^5$  & 10$^4$  & 10$^5$ \\
\hline            \hline
  \HH (para) &  0.899&  0.898      & 0.791  &0.789   & 0.681  &  0.678 \\
   \HH(para) ETL    &0.873  & 0.873 &   0.703 &      0.703     &0.475  &   0.475 \\
 \HH(ortho)        &   0.101& 0.102  &0.209  & 0.211&      0.319 &  0.322   \\
   \HH(ortho) ETL    &0.127  &0.127   &0.297 &   0.297     & 0.525    &   0.525     \\
 \HHHp (para)   &  2. 92(-8)     & 4.77(-9)       &   2.91(-8)  &    4.83(-9)    &  2.91(-8)  &  4.89(-9)  \\
 \HHHp (ortho)    &   1.45(-8)  &   2.37(-9)       &     1.62(-8)     &    2.71(-9)  &  1.76(-8)  &  2.96(-9)    \\
 \HHDp (para)  &  4.92(-11)    &   7.95(-12)    &    1.80(-11)      &   2.96(-12)          &  9.99(-12)  &   1.66(-12)    \\
 \HHDp (ortho)  & 6.19(-11)  &  1.01(-11)  &     3.10(-11)   &  5.12(-12)  & 2.08(-11)  & 3.48(-12) \\
 \HDDp (ortho)  & 7.04(-13) &   1.16(-13)  &      1.24(-13)      &  2.11(-14) & 4.40(-14)   &  7.51(-15)\\
 \HDDp (para)  &  1.68(-13)  &  2.72(-14)  &      3.28(-14)  &  5.39(-15) & 1.24(-14)   &  2.06(-15)     \\
 \CHHHp (para)  & 2.22(-11)   &     9.56(-13)  & 2.38(-11)           & 1.12(-12)& 2.59(-11) &  1.35(-12)\\
 \CHHHp (ortho)&    2.14  (-11) &  8.98(-13)  &    2.31(-11)    &  1.08(-12)        &  2.63(-11)& 1.45(-12\\
 \CHHDp           &     4.15(-12) &  2.79(-13)  & 4.94(-12) & 3.84(-13) & 3.79(-12) &2.76 (-13)\\
 HCN          &  7.74(-8)  &  1.58  (-8) &        7.17(-8)     & 1.24(-8)  &  6.67(-8)  & 9.96(-9)\\
 DCN/HCN &  1.4(-3) &    1.8(-3)     &     8.4(-4)   &      1.4(-3)  & 4.9(-4) &  8.5(-4)\\   
\HHCO &  1.75(-9)  &  4.02(-10) &   1.54(-9) & 3.71(-10)& 1.37(-9) & 3.39(-10)    \\
HDCO / $\HHCO$ & 0.085 & 0.12&  0.071& 0.10 & 0.043& 0.055  \\
   \HCOp &  7.38(-9)      &   3.13(-9)       &    8.51(-9)    &  3.76(-9)     &   9.56(-9)  &    4.36(-9)     \\
\DCOp / \HCOp & 1.4(-3) &   1.1(-3)       &    6.7(-4)   &  5.1(-4)  &  4.2(-4) & 3.1(-4) \\
    \NNHp &   1.39(-9)     &    9.03(-10)      &    1.65(-9)          &  1.10(-9)     &   1.92(-9)  &  1.28(-9)       \\
\NNDp / \NNHp &  1.4(-3) &  1.25(-3)  &    6.0(-4) &  5.3(-4)  & 3.4(-4)  &  3.1(-4)\\
\hline
    \end{tabular}
   \end{center}
   The elemental abundances are [D]/[H] = 1.6 $\times$ 10$^{-5}$, [C]/[H] = 7.0 $\times$ 10$^{-6}$, [O]/[H] =  2 $\times$  10$^{-5}$, [N]/[H]= 10$^{-5}$, [S]/[H] = 1.8 $\times$ 10$^{-7}$ and a representative metal [M]/[H] = 1.5 $\times$ 10$^{-8}$. Values in parentheses refer to power of 10.
 \end{table*}

\acknowledgement
This work  has greatly benefited from different aspects of  Oka's contributions in various areas.  
This research is partly based upon work at the Caltech Submillimeter Observatory, which is operated by the California Institute of Technology under cooperative agreement with the National Science Foundation (AST-0838261) and also on observations obtained with the IRAM-30 m telescope.
IRAM is supported by INSU/CNRS (France), MPG (Germany), and
IGN (Spain). Part of the work has been done when D. Lis was visiting LERMA as invited professor at the Physics Department of Ecole Normale Superieure. 

\bibliography{roueff}

\end{document}